\documentclass[aps,pra,twocolumn]{revtex4-1}

\usepackage{amsmath, amsfonts, amssymb, amsthm, newlfont, graphicx}
\usepackage{epstopdf}
\usepackage{color, pdfcolmk}
 \usepackage[normalem]{ulem}

\usepackage{natbib}

\newcommand{\pa}{\partial}

\newcommand{\be}{\begin{equation}}
\newcommand{\ee}{\end{equation}}

\begin{document}

\title{Amplitude death criteria for coupled complex Ginzburg-Landau systems}
\author{Robert A. Van Gorder}
\email{rvangorder@maths.otago.ac.nz}
\affiliation{Mathematical Institute, University of Oxford, Andrew Wiles Building, Radcliffe Observatory Quarter, Woodstock Road, Oxford, OX2 6GG, UK}
\affiliation{Department of Mathematics and Statistics, University of Otago,
P.O. Box 56, Dunedin 9054, NZ}
\author{Andrew L. Krause}
\affiliation{Mathematical Institute, University of Oxford, Andrew Wiles Building, Radcliffe Observatory Quarter, Woodstock Road, Oxford, OX2 6GG, UK}
\author{James A. Kwiecinski}
\affiliation{Mathematics, Mechanics, and Materials Unit, 
Okinawa Institute of Science and Technology, Okinawa, 904-0495, Japan}

\begin{abstract}
Amplitude death, which occurs in a system when one or more macroscopic wavefunctions collapse to zero, has been observed in mutually coupled solid-state lasers, analog circuits, and thermoacoustic oscillators, to name a few applications. While studies have considered amplitude death on oscillator systems and in externally forced complex Ginzburg-Landau systems, a route to amplitude death has not been studied in autonomous continuum systems. We derive simple analytic conditions for the onset of amplitude death of one macroscopic wavefunction in a system of two coupled complex Ginzburg-Landau equations with general nonlinear self- and cross-interaction terms. Our results give a more general theoretical underpinning for recent amplitude death results reported in the literature, and suggest an approach for tuning parameters in such systems so that they either permit or prohibit amplitude death of a wavefunction (depending on the application). Numerical simulation of the coupled complex Ginzburg-Landau equations, for examples including cubic, cubic-quintic, and saturable nonlinearities, is used to illustrate the analytical results.
\end{abstract}

\maketitle

\section{Introduction}
\label{intro}

Complex Ginzburg-Landau (GL) equations and generalizations to vector systems have been used to model nonlinear waves, second-order phase transitions, superconductivity, superfluidity, Bose-Einstein condensation, liquid crystals, and strings in field theory \cite{aranson2002world}, to name a few applications. Such equations exhibit a rich variety of dynamics, with pulses and fronts \cite{akhmediev2001pulsating,malomed2007solitary}, dissipative solitons \cite{akhmediev2005dissipative,skarka2006stability}, multi-solitons \cite{akhmediev1997multisoliton}, periodic solutions \cite{porubov1999exact}, vortex solutions \cite{chen1994shooting}, spiral waves \cite{biktasheva1998localized}, and optical rogue waves \cite{gibson2016optical} among the structures observed, while less ordered behavior such as spatiotemporal chaos and blow-up can be found as well \cite{chate1994spatiotemporal,shraiman1992spatiotemporal,weber1992stability}. While cubic nonlinearity is most common, there are other possibilities, the choice of which is motivated either by the application studied or the dynamics sought \cite{moores1993ginzburg,sugavanam2015ginzburg,dynamics}. 

Vector complex GL systems have been used to model phenomena ranging from electroconvection in planarly aligned nematic liquid crystals \cite{treiber1998coupled} to gas-less combustion fronts \cite{matkowsky1992coupled}. The interaction of different macroscopic wavefunctions becomes possible when considering vector complex GL systems. In the case where the squared-modulus of each wavefunction features into each equation, the cross-phase modulation (XPM) parameters may vary, meaning that the form of the nonlinear term in each equation will differ \cite{malomed2012unstaggered,sakaguchi1995phase}. Phase instabilities in vector GL equations have been studied \cite{san1995phase}, and such instabilities and spatiotemporal chaos appear quite commonly in complex GL systems \cite{dynamics}. 

The amplitude death of one macroscopic wavefunction under asymmetric dynamics which favor the propagation of the other wavefunction has attracted recent interest in mathematical physics and the study of nonlinear waves. Amplitude death or partial amplitude death is more commonly seen in literature on network or lattice equations such as oscillator systems \cite{astakhov2011peculiarities, banerjee2015mean, dodla2004phase, ermentrout1990oscillator, koseska2013oscillation, liu2012effects, mehta2006death, mirollo1990amplitude, nakao2009diffusion, nakao2014complex, resmi2011general, saxena2012amplitude}. For the continuum PDE case there is far less work. Existing work considers control of such dynamics in reaction-diffusion systems \cite{stich2007control,teki2017amplitude}. In particular, \cite{stich2007control} use a mixed local and global control to obtain amplitude death solutions (among other kinds of solutions), while \cite{teki2017amplitude} shows that amplitude death can occur in a pair of one-dimensional cubic complex GL systems coupled by a controller using diffusive connections. The analytical results of \cite{teki2017amplitude} reveal that amplitude death never occurs in a pair of identical complex cubic GL systems, hence one does not have amplitude death in vector complex GL systems where the reaction kinetic parameters are equal in each equation. Synchronization in complex cubic GL systems with asymmetries was earlier considered in \cite{zhou2006synchronization}, and while a variety of dynamics were explored, it does not appear the role of asymmetry due to XPM in amplitude death was explored. In the case of saturable nonlinearity, it was recently shown \cite{dynamics} that amplitude death emerges from complex GL systems when XPM parameters are large enough relative to self-interaction terms without any form of delay coupling or control. The existence of amplitude death has been observed in mutually coupled solid-state lasers \cite{wei2007amplitude}, analog circuits \cite{suresh2014experimental}, and thermoacoustic oscillators \cite{biwa2015amplitude}, and this suggests that amplitude death in GL systems merits further study, perhaps in the context of BEC and chemical systems governed by complex GL systems involving multiple scalar wavefunctions.

Motivated by these findings, we obtain an analytical criteria for the amplitude death of wavefunctions in complex GL systems with generic kinetics. We consider a general coupled complex GL system taking the form 
\begin{align}
\frac{\pa u}{\pa t} & = (\epsilon + i\hat{\epsilon}) u + \left( a + i\hat{a}\right)\nabla^2 u - \left( b + i\hat{b}\right) f\left( |u|^2, |v|^2\right)u\,, \label{sgl1} \\
\frac{\pa v}{\pa t} & = (\epsilon + i\hat{\epsilon}) v + \left( a + i\hat{a}\right)\nabla^2 v - \left( b + i\hat{b}\right) g\left( |u|^2, |v|^2\right)v\,,\label{sgl2}
\end{align}
where $u,v\in \mathbb{C}$ are complex wavefunctions defined on Euclidean space $\mathbb{R}^m$ for $m\geq 1$, $t\in [0,\infty)$, and all of $\epsilon, \hat{\epsilon}, a, \hat{a}, b, \hat{b}$ are real-valued parameters. We shall always assume $a,b>0$. Here $\nabla^2 = \frac{\partial^2}{\partial x_1^2} + \cdots + \frac{\partial^2}{\partial x_m^2}$ denotes the Laplacian operator in $m$ spatial coordinates. We consider $f$ and $g$ as general nonlinear functions satisfying $f(0,0)=g(0,0)=0$, and take $f$ and $g$ to be strictly monotone increasing in both arguments (in agreement with most applications). Both restrictions are motivated by commonly used and physically relevant forms of $f$ and $g$ employed in the literature. Noting that there are a wide variety of experimental approaches for realizing XPM in nonlinear optics \cite{petrosyan2004magneto,matsko2005optical,lo2011electromagnetically}, quantum information \cite{paternostro2003generation,perrella2013high}, and many other applications, we keep $f$ and $g$ general in our analysis, so that a user may insert the specific choice of $f$ and $g$ relevant to their system of interest.

We obtain generic conditions leading to amplitude death of one wavefunction in \eqref{sgl1}-\eqref{sgl2}, and these criteria are derived and presented in Sec. \ref{sec3.1}. In Sec. \ref{numerics} we employ numerical simulations to illustrate the analytical results for particular $f$ and $g$ commonly used in the literature, including (i) $f(A,B) = A + \alpha_1 B$ and $g(A,B) = \alpha_2 A + B$ for which we obtain the standard cubic complex GL system (where $\alpha_1$ and $\alpha_2$ are XPM parameters); (ii) $f(A,B)$ and $g(A,B)$ are quadratic functions, in which case we obtain a cubic-quintic GL system; and (iii)  $f(A,B) = \frac{A + \alpha_1 B}{\left( 1+ \mu (A+B)^n\right)^{1/n}}$, $g(A,B) = \frac{\alpha_2 A + B}{\left( 1+ \mu (A+B)^n\right)^{1/n}}$, in which case we obtain a complex GL system with saturable nonlinearity, appearing in the study of nonlinear optics in saturable media \cite{tikhonenko1996three,soto1991stability,yulin2008discrete,raja2010modulational}. Finally, in Sec. \ref{disc}, we summarize our results and give concluding remarks. We also compare our results with reductions to simpler cases, to demonstrate why amplitude death is not found in simpler systems, such as vector NLS equations, and compare our results to a family of dark solitons which give decay of one wavefunction, but only as $t\rightarrow \infty$.

\section{Criteria for amplitude death}\label{sec3.1}
In order to study amplitude death of wavefunctions, we shall begin by constructing an approximation to the attractor $\mathcal{A}\subset \mathbb{C}^2$ for solutions $(u,v)\in\mathbb{C}^2$ (such that all trajectories $(u,v)\in \mathcal{A}$ as $t\rightarrow \infty$). We shall also make the simplifying assumption that there is one dominant spatial mode for sake of analytical tractability, with this corresponding to the minimal spatial wavenumber. Assuming that arbitrarily small modes are excited on the plane (say, due to random noise, or similar small perturbations), we then take the wavenumbers corresponding to dominant modes to be arbitrarily small, with zero wavenumbers giving the tightest bounds which guarantee the amplitude death. More rigorous bounds on attractors for complex GL equations have been considered in \cite{mielke1997complex,mielke1998bounds}.

To begin, we consider a solution of the form
\be \label{tranf}
u = e^{i\mathbf{k}_u\cdot \mathbf{x}}U(t) \quad \text{and} \quad v = e^{i\mathbf{k}_v\cdot \mathbf{x}}V(t)\,.
\ee
Such a solution will give the dynamics corresponding to a single wavenumber. The transformation \eqref{tranf} puts \eqref{sgl1}-\eqref{sgl2} into the form
\be 
\frac{dU}{dt}  = (\epsilon + i\hat{\epsilon}) U - \left( a + i\hat{a}\right)|\mathbf{k}_u|^2 U - \left( b + i\hat{b}\right) f\left( |U|^2, |V|^2\right)U\,, \label{T1}
\ee 
\be 
\frac{dV}{dt}  = (\epsilon + i\hat{\epsilon}) V - \left( a + i\hat{a}\right)|\mathbf{k}_v|^2 V - \left( b + i\hat{b}\right) g\left( |U|^2, |V|^2\right)V\,.\label{T2} 
\ee
Consider $U(t) = \rho_u\exp(i\theta_u)$, $V(t)=\rho_v\exp(i\theta_v)$, which puts \eqref{T1}-\eqref{T2} into the form
\begin{align}
\frac{d\rho_u}{dt} & = \left(\epsilon - a|\mathbf{k}_u|^2\right)\rho_u - b f\left( \rho_u^2, \rho_v^2\right)\rho_u\,,\label{rhou} \\
\rho_u\frac{d\theta_u}{dt} & = \left(\hat{\epsilon}- \hat{a}|\mathbf{k}_u|^2\right) \rho_u - \hat{b} f\left( \rho_u^2, \rho_v^2\right)\rho_u \,,\\
\frac{d\rho_v}{dt} & = \left(\epsilon - a|\mathbf{k}_v|^2\right)\rho_v - b g\left( \rho_u^2, \rho_v^2\right)\rho_v\,, \label{rhov}\\ 
\rho_v\frac{d\theta_v}{dt} & = \left( \hat{\epsilon} - \hat{a}|\mathbf{k}_v|^2 \right)\rho_v  - \hat{b}g\left( \rho_u^2, \rho_v^2\right)\rho_v\,.
\end{align}
The dynamics for $\theta_u$ and $\theta_v$ decouple and can be found immediately once the dynamics of $\rho_u$ and $\rho_v$ are known. In order for a non-trivial asymptotic solution, we should have $\epsilon > a \max\left\lbrace |\mathbf{k}_u|^2 ,  |\mathbf{k}_v|^2 \right\rbrace$, otherwise the zero state $(u,v)=(0,0)$ is locally stable. If the wavenumber perturbations are taken arbitrarily small, then this condition reduces to $\epsilon >0$. This means that the attractor for the amplitudes, $\mathcal{A}'\subset [0,\infty)^2$, is bounded away from zero.

Writing $\rho_u = r \cos(\phi)$, $\rho_v = r \sin(\phi)$, where $r\geq 0$ and $\phi \in \left[ 0, \frac{\pi}{2}\right]$, we have that \eqref{rhou} and \eqref{rhov} are equivalent to
\be \label{rode}\begin{aligned}
\frac{dr}{dt} & = \left\lbrace \epsilon - a \left( |\mathbf{k})u|^2 \cos^2(\phi) + |\mathbf{k}_v|^2 \sin^2(\phi) \right)  \right. \\
& \qquad - b f\left( r^2\cos^2(\phi), r^2\sin^2(\phi)\right)\cos^2(\phi)\\
& \qquad \left. - b g\left( r^2\cos^2(\phi), r^2\sin^2(\phi)\right)\sin^2(\phi)  \right\rbrace r\,,
\end{aligned}\ee
\be \begin{aligned}\label{phiode}
\frac{d\phi}{dt} & = \left\lbrace a\left( |\mathbf{k}_u|^2 - |\mathbf{k}_v|^2\right) + b f\left( r^2\cos^2(\phi), r^2\sin^2(\phi)\right) \right. \\
& \qquad \left. - b g\left( r^2\cos^2(\phi), r^2\sin^2(\phi)\right)\right\rbrace \sin(\phi)\cos(\phi)\,.
\end{aligned}\ee
Assuming $\epsilon > a \max\left\lbrace |\mathbf{k}_u|^2 ,  |\mathbf{k}_v|^2 \right\rbrace$, then the attractor consists of a curve $r=Q(\phi)$ such that $\frac{dr}{dt}\equiv 0$ on $r=Q(\phi)$.

To ensure that the attractor $\mathcal{A}'$ is bounded away from infinity, we require $\frac{dr}{dt}<0$ for large $r$. From the form of \eqref{rode}, we see that this condition is equivalent to
\be \begin{aligned}
\epsilon & - a \left( |\mathbf{k}_u|^2 \cos^2(\phi) + |\mathbf{k}_v|^2 \sin^2(\phi) \right) \\
& \qquad - b f\left( r^2\cos^2(\phi), r^2\sin^2(\phi)\right)\cos^2(\phi) \\
& \qquad - b g\left( r^2\cos^2(\phi), r^2\sin^2(\phi)\right)\sin^2(\phi)  < 0
\end{aligned}\ee
as $r\rightarrow \infty$, for all $\phi \in \left[ 0, \frac{\pi}{2}\right]$. The least stable state is for arbitrarily small wavenumbers, so we require $\epsilon < bL$, where
\be 
L = \min_{\chi\in [0,1]} \lim_{s\rightarrow \infty}f\left( \chi s, (1-\chi)s\right)\chi + g\left( \chi s, (1-\chi)s\right)(1-\chi)\,.
\ee 
That such a limit should be positive follows from monotonicity of $f$ and $g$. Note that we may have $L = \infty$, which just implies that the dynamics are bounded for all $\epsilon >0$. 

We have found that the conditions for an attractor $\mathcal{A}'$ to be bounded away from both infinity and zero is that $\epsilon$ satisfy the bound $0 < \epsilon < bL$. This, in turn, implies that $0< |u|^2+|v|^2 <\infty$ under these parameter restrictions. The result holds for any fixed wavenumbers $\mathbf{k}_u$ and $\mathbf{k}_v$, and since the wavenumber will only result in a shift in the $\epsilon$ parameter bound ($\epsilon > a \max\left\lbrace |\mathbf{k}_u|^2 ,  |\mathbf{k}_v|^2 \right\rbrace$), we shall be justified in taking $\mathbf{k}_u = \mathbf{k}_v = \mathbf{0}$ when convenient. If one is interested in small finite domains for which there are minimal wavenumbers, one may instead replace the obtained bound with $a \max\left\lbrace |\mathbf{k}_{u,\min}|^2 ,  |\mathbf{k}_{v,\min}|^2 \right\rbrace < \epsilon < b L$, where $\mathbf{k}_{u,\min}$ and $\mathbf{k}_{v,\min}$ denote the respective minimal wavenumbers in such a case.

The function $Q(\phi)$ is implicitly defined by the relation
\be \label{implicit}\begin{aligned}
 \epsilon & -  a \left( |\mathbf{k})u|^2 \cos^2(\phi) + |\mathbf{k}_v|^2 \sin^2(\phi) \right)  \\
 &  - b f\left( Q(\phi)^2\cos^2(\phi), Q(\phi)^2\sin^2(\phi)\right)\cos^2(\phi) \\
 &  - b g\left( Q(\phi)^2\cos^2(\phi), Q(\phi)^2\sin^2(\phi)\right)\sin^2(\phi)   = 0\,,
\end{aligned}\ee
and we use \eqref{implicit} to find $f\left(Q(0)^2,0\right) = \frac{\epsilon - a|\mathbf{k}_u|^2}{b}$, $g\left(0,Q(\pi/2)^2\right) = \frac{\epsilon - a|\mathbf{k}_v|^2}{b}$. Define $f_u^{-1}$ to be the inverse function of $f$ with respect to the first argument when the second argument is zero. That is to say, $f\left( f_u^{-1}(y),0\right)=y$. As we assume $f$ is strictly monotone increasing in both arguments, this inverse function will be defined globally. Then, we may write
\be 
Q(0) = \sqrt{f_u^{-1}\left( \frac{\epsilon - a|\mathbf{k}_u|^2}{b}\right)}\,.
\ee
Similarly, defining $g_v^{-1}$ to be the inverse function of $g$ with respect to the second argument when the first argument is zero, viz., $g\left( 0, g_v^{-1}(y) \right)=y$, we have that
\be 
Q\left( \frac{\pi}{2}\right) = \sqrt{g_v^{-1}\left( \frac{\epsilon - a|\mathbf{k}_v|^2}{b}\right)}\,.
\ee

Making use of $r=Q(\phi)$ on the attractor, then $\frac{d\phi}{dt} = \Phi(\phi)$ on the attractor, where from \eqref{phiode} we find
\be \begin{aligned}
\Phi(\phi) & = \left\lbrace a\left( |\mathbf{k}_u|^2 - |\mathbf{k}_v|^2\right) \right.\\
& \quad + b f\left( Q(\phi)^2\cos^2(\phi), Q(\phi)^2\sin^2(\phi)\right)  \\
& \quad \left. - b g\left( Q(\phi)^2\cos^2(\phi), Q(\phi)^2\sin^2(\phi)\right)\right\rbrace \sin(\phi)\cos(\phi)\,.
\end{aligned}\ee
As the attractor corresponds to $r=Q(\phi)$ then it is sufficient to determine the stability of a particular state corresponding to $(r_*,\phi_*) = (Q(\phi_*),\phi_*)$ from the sign of $\Phi'(\phi_*)$. 

\subsection{Amplitude death of $u$}
The case of amplitude death of the wavefunction $u$ corresponds to $\phi = \frac{\pi}{2}$, and we find that the state $\phi = \frac{\pi}{2}$ is locally stable provided that $\Phi'\left( \frac{\pi}{2}\right) <0$, i.e.,
\be 
\epsilon - a |\mathbf{k}_u|^2 - b f \left( 0, g_v^{-1}\left( \frac{\epsilon - a|\mathbf{k}_v|^2}{b} \right)\right) < 0\,,
\ee
provided that $a \max\left\lbrace |\mathbf{k}_{u}|^2 ,  |\mathbf{k}_{v}|^2 \right\rbrace < \epsilon < b L$.
If we consider dynamics on $\mathbb{R}^m$, $m \geq 1$, then arbitrarily small wavenumbers will be excited by arbitrary perturbations (such as random noise), so we take $\mathbf{k}_u = \mathbf{k}_v =0$. In this case, the amplitude death state for $u$ is locally stable if 
\be \label{cc1}
\frac{\epsilon}{b} <  f \left( 0, g_v^{-1}\left( \frac{\epsilon}{b} \right)\right)  \,,
\ee
provided $0 < \epsilon < bL$.

\subsection{Amplitude death of $v$}
The case of amplitude death of the wavefunction $v$ corresponds to $\phi = 0$, and we find that the state $\phi = 0$ is locally stable provided that $\Phi'\left( 0\right) < 0$, i.e.,
\be 
\epsilon - a |\mathbf{k}_v|^2 - b g \left( f_u^{-1}\left( \frac{\epsilon - a|\mathbf{k}_u|^2}{b}\right), 0\right) < 0\,,
\ee
provided that $a \max\left\lbrace |\mathbf{k}_{u}|^2 ,  |\mathbf{k}_{v}|^2 \right\rbrace < \epsilon < b L$.
If we consider dynamics on $\mathbb{R}^m$, $m \geq 1$, then arbitrarily small wavenumbers will be excited by arbitrary perturbations, so we take $\mathbf{k}_u = \mathbf{k}_v =0$. In this case, the amplitude death state for $v$ is locally stable if 
\be \label{cc2}
\frac{\epsilon}{b} <  g \left( f_u^{-1}\left( \frac{\epsilon}{b} \right), 0\right) \,,
\ee
provided $0 < \epsilon < bL$.

\begin{figure*}
\begin{center}

\includegraphics[width=0.4\textwidth]{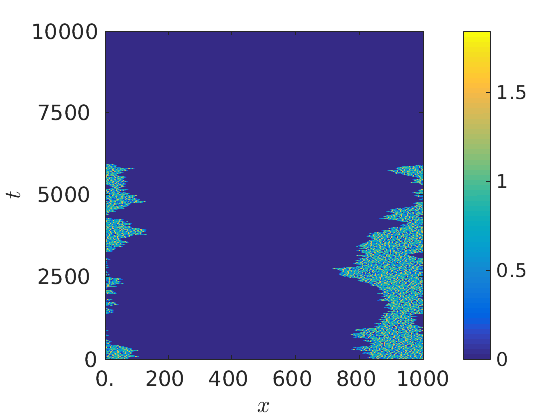}
\includegraphics[width=0.4\textwidth]{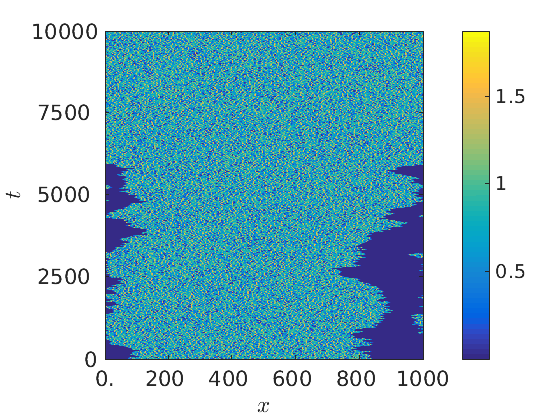}

(a) \hspace{7cm} (b)
\vspace{-0.1in}
\caption{Plot of (a) $|u|^2$, (b) $|v|^2$ with cubic interaction terms, $\epsilon=1$, $\hat{a}=5$, $\hat{b}=-1$, $a=1$, $b=1$, and $\hat{\epsilon}=0$ with periodic boundary conditions on a domain of length $L=1000$. We take $\alpha_1=\alpha_2=10$.\label{Fig0}
}
\end{center}
\end{figure*}

\subsection{Summarizing remarks}
For each case, we note that the local conditions on amplitude death rely on both $f$ and $g$. In particular, the relative strength of both functions in the $|u|^2$ or $|v|^2$ argument (as measured via function composition) will determine if the amplitude death of the wavefunction for the other argument, $|v|^2$ or $|u|^2$, is stable. As the results are local, solutions corresponding to initial conditions far from the region of the attractor permitting amplitude death may result in dynamics which persist over time. As we shall later show numerically, the excitation of a number of wavenumbers over time often results in spatiotemporal chaos, and this can permit solutions to enter the amplitude death regime eventually, after which point one of the two amplitudes will decay. Therefore, initially structured solutions may degenerate into spatiotemporal chaos, which may later result in one solution undergoing amplitude death, provided that the parameter restrictions we derived here are obeyed.

\begin{figure*}
\begin{center}

\includegraphics[width=0.4\textwidth]{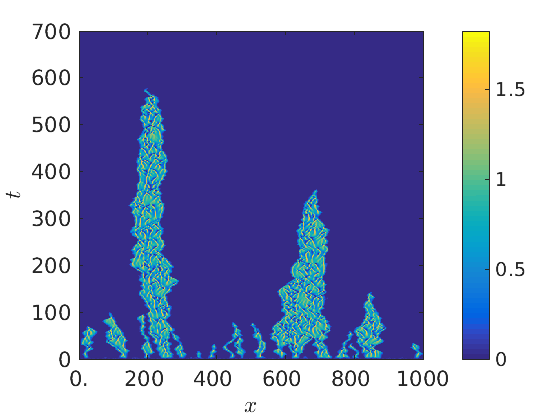}
\includegraphics[width=0.4\textwidth]{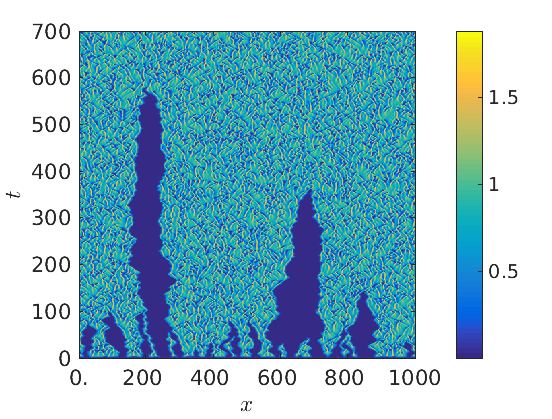}

(a) \hspace{7cm} (b)
\vspace{-0.1in}
\caption{Plot of (a) $|u|^2$, (b) $|v|^2$ with cubic interaction terms, $\epsilon=1$, $\hat{a}=5$, $\hat{b}=-1$, $a=1$, $b=1$, and $\hat{\epsilon}=0$ with periodic boundary conditions on a domain of length $L=1000$. We consider the asymmetric case $\alpha_1=10$ and $\alpha_2=9.5$.\label{Fig1}
}
\end{center}
\end{figure*}

\section{Examples of amplitude death}\label{numerics}
We now provide specific examples of amplitude death for different interaction functions $f$ and $g$. We numerically solve \eqref{sgl1}-\eqref{sgl2} by discretizing the spatial derivatives via centered finite differences, and using the Matlab function `ode45' which implements an adaptive Runge-Kutta scheme in time. We take small random initial data for the real and complex parts of each wavefunction by sampling from a spatially-distributed uniform process with values in $(0,1)$. Simulations with Gaussian initial data gave qualitatively comparable behavior over many realizations. We are interested in the behavior of the GL system on infinite domains, so to approximate this we discretized a spatial domain of length $L=10^3$ using $2\times 10^3$ nodal points, and imposed periodic conditions unless specified otherwise. We used absolute and relative error tolerances of $10^{-9}$, and performed convergence checks in the discretization scheme, alongside an independent implementation using the finite element software COMSOL to ensure convergence.

\begin{figure*}
\begin{center}
\includegraphics[width=0.4\textwidth]{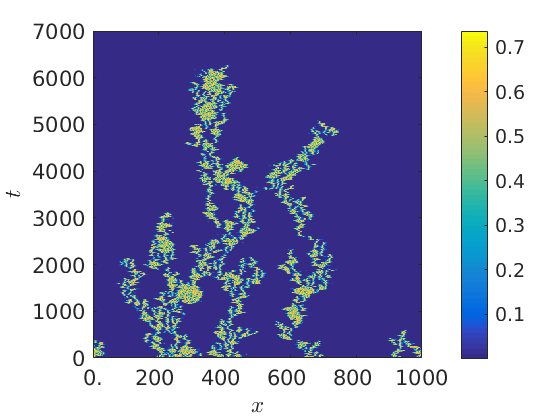}
\includegraphics[width=0.4\textwidth]{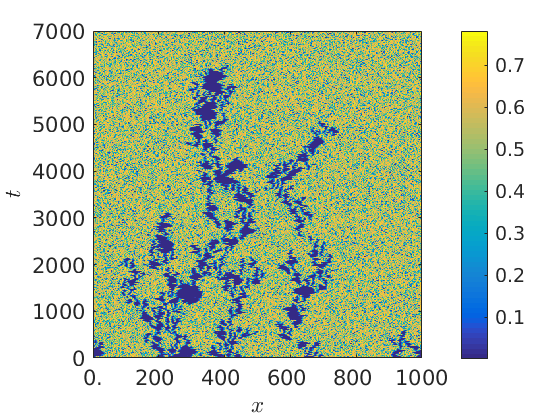}

(a) \hspace{7cm} (b)
\vspace{-0.1in}
\caption{Plots of $|u|^2$ in (a) and $|v|^2$ in (b) with cubic-quintic interaction terms, $\epsilon=1$, $\hat{a}=2$, $\hat{b}= -1$, $a=1$, $b=1$, $\alpha_1=\alpha_2=1$ and $\hat{\epsilon}=0$ with periodic boundary conditions on a domain of length $L=1000$. For the quintic terms, we take $\beta_{11}=1.2$, $\beta_{12}=5$, $\beta_{21}=5$, and $\beta_{22}=1$. }\label{Fig2}
\end{center}
\end{figure*}

\paragraph{Cubic kinetics}
Consider $f(A,B)=A+\alpha_1 B$, $g(A,B)= \alpha_2 A + B$, which gives the complex cubic GL system. Then, $f_u^{-1}(y)=y$, $g_v^{-1}(y)=y$, and $\Phi'(0) = (1-\alpha_2)\epsilon$, $\Phi'\left( \frac{\pi}{2}\right) = (1-\alpha_1)\epsilon$. This suggests that amplitude death of $u$ is locally stable when $\alpha_1 >1$, while amplitude death of $v$ is locally stable when $\alpha_2 >1$. In Fig. \ref{Fig0} we take $\alpha_1 = \alpha_2 =10$, and observe global amplitude death of one of the wavefunctions (in this case, $u$) over a long finite time scale. In Fig. \ref{Fig1}, we take $\alpha_1=10$ and $\alpha_2=9.5$ and demonstrate amplitude death of $u$ over a shorter time frame due to this asymmetry. We remark that such spatial separation of the wavefunctions occurs over wide ranges of parameter space, so long as $\alpha_j > 1$ for at least one $j=1,2$. Within the region where each wavefunction is non-zero, we observe spatiotemporal chaos, which is common in complex GL systems.

Note that the system \eqref{sgl1}-\eqref{sgl2} with cubic nonlinearity may admit exact solutions in some parameter regimes, for which the system becomes conditionally integrable. For rather restrictive parameter sets, one may find a family of dark solitons \cite{kivshar1998dark,kivshar1993vector}, with one of $u$ or $v$ behaving like $1-\tanh(\mathbf{k}\cdot \mathbf{x} + ct)$ with wavenumber vector $\mathbf{k}$ and wavespeed $c>0$. In this case, the soliton solution will exhibit behavior akin to amplitude death for very large time (e.g., as $t \rightarrow \infty$). In contrast, the amplitude death we study often occurs in finite time, and occurs more generally in the non-integrable regime, which is a far larger subset of parameter space.

\paragraph{Cubic-quintic kinetics}
Consider $f(A,B)=A+\alpha_1 B + \beta_{11}A^2 + \beta_{12}B^2$, $g(A,B)= \alpha_2 A + B + \beta_{21}A^2 + \beta_{22}B^2$. We find that
$$
f_u^{-1}(y) = \frac{\sqrt{1+4\beta_{11}y} -1}{2}, \quad g_v^{-1}(y) = \frac{\sqrt{1+4\beta_{22}y} -1}{2}\,.
$$
Amplitude death of the wavefunction $u$ is then locally stable provided
$$
\frac{\epsilon}{b} < \alpha_1 g_v^{-1}\left( \frac{\epsilon}{b}\right) + \beta_{12} \left( g_v^{-1}\left( \frac{\epsilon}{b}\right)\right)^2\,,
$$
which gives the parameter restriction
$$ 
\frac{\alpha_1 - \beta_{12}}{2}\left\lbrace \sqrt{1 + \frac{2\beta_{22}\epsilon}{b}} -1 \right\rbrace + (\beta_{12}\beta_{22} -1)\frac{\epsilon}{b} >0\,.
$$
Likewise, amplitude death of the wavefunction $v$ is locally stable provided
$$ 
\frac{\epsilon}{b} < \alpha_2 f_u^{-1}\left( \frac{\epsilon}{b}\right) + \beta_{21} \left( f_u^{-1}\left( \frac{\epsilon}{b}\right)\right)^2\,,
$$
which gives the parameter restriction
$$
\frac{\alpha_2 - \beta_{21}}{2}\left\lbrace \sqrt{1 + \frac{2\beta_{11}\epsilon}{b}} -1 \right\rbrace + (\beta_{11}\beta_{21} -1)\frac{\epsilon}{b} >0\,.
$$
In Fig. \ref{Fig2} we see that a slight asymmetry in the quintic-order terms leads to one of the wavefunctions exhibiting  amplitude death.

\begin{figure*}
\centering
\includegraphics[width=0.4\textwidth]{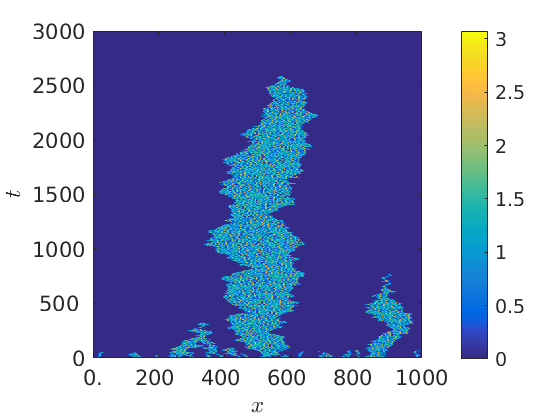}
\includegraphics[width=0.4\textwidth]{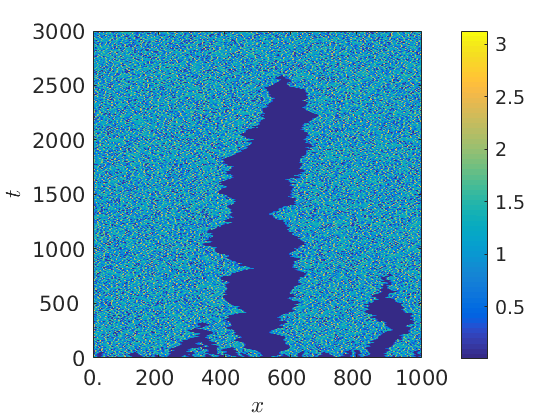}

(a) \hspace{7cm} (b)
\vspace{-0.1in}
\caption{Plots of $|u|^2$ in (a) and $|v|^2$ in (b) with saturable interaction terms and parameters (and initial data) identical to that of Figure \ref{Fig1}. Note the change in timescales. We fix $\mu=0.3$.}\label{Fig3}
\end{figure*}

\paragraph{Saturable kinetics}
Given the saturable kinetics
$$
f(A,B) = \frac{A + \alpha_1 B}{\left( 1+ \mu (A+B)^n\right)^{1/n}}, 
$$
$$ 
g(A,B) = \frac{\alpha_2 A + B}{\left( 1+ \mu (A+B)^n\right)^{1/n}},
$$
with saturation parameter $\mu >0$, we have 
$$ 
 f_u^{-1}(y) = g_v^{-1}(y) = \frac{y}{(1-\mu y^n)^{1/n}}\,,
$$
from which we again find $\Phi'(0) = (1-\alpha_2)\epsilon$, $\Phi'\left( \frac{\pi}{2}\right) = (1-\alpha_1)\epsilon$, giving the same criteria as found for the cubic kinetics. In Fig. \ref{Fig3}, we give numerical simulations using such kinetics when $n=1$. In particular we use the same parameters, as well as identically the same realization of initial data, as was used in Fig. \ref{Fig1}. We observe that the saturation parameter, $\mu$, does not appear to change the long-time behavior, but substantially changes the transient dynamics of the system. In particular, in Fig. \ref{Fig3} we show a similar amplitude death in the wavefunction $u$, but this occurs over a time scale roughly five times longer than the corresponding dynamics for cubic kinetics in Fig. \ref{Fig1}. For larger values of the saturation parameter, the system is pushed out of the parameter set corresponding to spatiotemporal chaos, and the system exhibits banded patterns which become stationary (in modulus) solutions. We remark that these dynamics appear robust to various realizations of initial data; see \cite{dynamics} for additional simulations of saturable complex Ginzburg-Laundau equations.

\section{Discussion}\label{disc}
Starting with a general complex Ginzburg-Landau system \eqref{sgl1}-\eqref{sgl2}, we were able to obtain generic local conditions for the amplitude death of a single wavefunction. While local in nature, the results are analytical, and can easily be verified via numerical simulations, which suggests that the global dynamics behave akin to what the local theory suggests. The analytical conditions on amplitude death of either wave function suggest that asymmetry in the nonlinear self- and cross-interaction terms will be responsible for this amplitude death. This was suggested by the results of \cite{teki2017amplitude}, as in that paper symmetric coupled GL systems were considered and the only route to amplitude death was due to a very specific controller which enters into the GL system as a forcing function. Our results give a more systematic view of the asymmetry-induced amplitude death previously reported for very specific cases of nonlinearity in \cite{dynamics}. Indeed, as our results suggest, the control and intentional destruction of certain wavefunctions is possible by appropriately tuning the ratio of cross-phase modulation to self-interaction terms, and such a controller might be realized by using time-dependent parameter values (see, for instance, \cite{baines2018soliton}) in order to force or prohibit amplitude death depending on the needs of the application. Thus, the asymmetry required for amplitude death may be exploited in nonlinear optics to reproduce results akin to ours in laboratory experiments, as the interaction between self- and cross-phase modulation has already been exploited experimentally in a number of studies  \cite{agrawal1987modulation,chiang1994cross,genty2004effect,hofer1991mode,hong1996femtosecond}. One could, for instance, imagine a modification of \cite{wei2007amplitude} for mutually coupled solid-state lasers which accounts for asymmetry.

The asymmetry inducing amplitude death cannot occur in scalar complex GL equations, and hence a system of such equations is required. Additionally, we should remark that in the case of nonlinear Schr\"odinger (NLS) systems, we necessarily have that $b=0$ (among other parameter restrictions), which puts \eqref{rhou} and \eqref{rhov} into the form $\frac{d\rho_{u,v}}{dt} = \left( \epsilon - a |\mathbf{k}_{u,v}|^2\right) \rho_{u,v}$. Depending on $\epsilon$, the dynamics tend to zero or infinity. Therefore, the amplitude death of a single wavefunction is not observed in coupled NLS systems involving two equations, and hence requires the dissipative dynamics inherent in complex GL systems.

Our results can be extended to systems with three or more wavefunctions. Indeed, there will possibly be many more routes to amplitude death in such high-order equations, and it may be interesting to study situations where a certain subset of the wavefunctions will undergo amplitude death; for a general system of $N$ complex GL equations, it may be possible to have up to $N-1$ wavefunctions exhibit amplitude death. This may have application to signal processing or the transmission of desired pulses/destruction of undesired pulses in fiber optics. Hence, the selection of amplitude death regimes in high-order complex GL systems is worthy of further consideration.

%

\end{document}